\documentstyle[aps,prl,psfig,floats]{revtex}
\begin{document}

\renewcommand{\topfraction}{.9}
\renewcommand{\bottomfraction}{.1}
\renewcommand{\textfraction}{.1}
\renewcommand{\floatpagefraction}{1}
\renewcommand{\dbltopfraction}{1}
\renewcommand{\dblfloatpagefraction}{1}

\twocolumn[\hsize\textwidth\columnwidth\hsize\csname %
@twocolumnfalse\endcsname

\title{An Empirical Pair-breaking Picture for the Non-linear Meissner Effect in $d$-wave Superconductors}

\author{J.R. Cooper}
\address{IRC in Superconductivity and Physics Department,\\ University of Cambridge, Madingley Road, Cambridge, CB3 OHE, U.K.}
\date{\today}
\maketitle {}
\begin{abstract}
A empirical model is proposed for the observed variation of the London
penetration depth of cuprate superconductors with applied magnetic field
$H$ below the lower critical field $H_{c1}$.   It is suggested that the
Doppler shift of the quasi-particle energies near the $d$-wave nodes
could cause pair-breaking and that this could be responsible for the
non-linear Meissner effect (NLME) investigated in recent experiments.
This  picture gives a somewhat better description of the observed
temperature and field dependencies than the original theory proposed by
Yip and Sauls.  A calculation based on weak-coupling BCS theory suggests
that when there is a finite supercurrent, the state for which pairs are
broken in a small angular range around the nodes can have a lower free
energy than the $d_{x^2-y^2}$ state.
\end{abstract}


]

\section{INTRODUCTION}
There is now a large body of experimental evidence \cite{harlingen,tsuei,tsueiel} showing that the order parameter
of all cuprate superconductors has unconventional $d_{x^2-y^2}$ symmetry. The superconducting gap
parameter\cite{waldram} $\Delta_{\vec{k}}$ varies as $\Delta\cos{2\theta}$ in the $k_{x},k_{y}$ plane of reciprocal
($\vec{k}$) space, where $\theta = \arctan{k_{y}/k_{x}}$, giving rise to line nodes at $45^{o}$ to the x and y
symmetry axes, at which the gap goes to zero and the order parameter changes sign. In a seminal paper, Yip and Sauls
predicted\cite{Yip} that there should be a non-linear Meissner effect(NLME) arising from the anomalous back flow of
Bogoliubov quasi-particles near these nodes. They showed that at low temperatures $(T)$ the London penetration depth
$\lambda$ should be increased significantly by small applied fields $H$ below the lower critical field $H_{c1}$, and
that there should also be non-linear effects in the transverse magnetisation\cite{Yip}. The NLME has been studied
theoretically by several groups \cite{Yip,Xu,scalapino,joynt,stojkovic} who all agree that $\lambda$ should increase
as $H^{1}$ at very low temperatures, when the Doppler or energy shift $d$ in the quasiparticle spectrum caused by the
flow of supercurrent is larger that $k_{B}T$. For YBa$_{2}$Cu$_{3}$O$_{6+x}$ crystals this condition is only
satisfied below approximately 2 K for fields of order $0.5H_{c1}$ or 300 Gauss. At higher temperatures, where
$k_{B}T \gg d$, the theory predicts that the non-linear effect in $\lambda$ should be much smaller and vary as $H^{2}$.
In contrast to these predictions, several experimental studies\cite{Maeda,Maedarev,Maedael,carrington,bidpre} show
that $\lambda$ increases as $H^{1}$ over most of the temperature range studied (typically 1.4 to 80 K), and that
d$\lambda$/d$H$ becomes larger at higher temperature. Moreover the expected anisotropy in $\delta\lambda (H)$, an
increase by a factor $\sqrt{2}$ when $H$ is applied along the nodal directions, and the predicted non-linear
transverse magnetic moment have not been observed experimentally\cite{bidinosti,valls}. One possible explanation for
these discrepancies is the occurrence of non-local effects\cite{li} at low $T$. However in one of the commonly used
experimental geometries, where the currents and fields are in the CuO$_{2}$ planes, these effects are governed by the
c-axis coherence length and would generally be small, although it has been argued \cite{li} that they would still
be significant in YBCO.

It is generally agreed that the experiments are difficult and not very reproducible, and it has been suggested that
the lack of reproducibility could be caused by trapped flux\cite{carrington} or the sharp edges of the
crystals\cite{bidinosti}.  My original motivation was to see whether the unusual $H^{1}$ dependence of $\lambda$
mentioned above and perhaps some of the apparent lack of reproducibility could be linked with other physical
properties which are extremely dependent on the hole concentration ($p$). For example the normal state pseudogap
develops suddenly when $p$ is less than 0.19 holes per planar Cu atom\cite{loram1,jwlhouston}, and this could
conceivably lead to unusual effects in $\lambda (H)$.  The thermoelectric power is often large \cite{obertelli},
implying substantial electron-hole asymmetry which might also affect the NLME. However  simple model calculations
\cite{cooperunp} based on standard theory seem to indicate  that none of these mechanisms
can cause  $\lambda$ to increase linearly with $H$ at higher temperatures.

As discussed in the present paper, a possible explanation of the experimental findings is that Cooper pairs near the
nodes of any  superconductor with higher order angular momentum pairing are actually broken when the energy difference
between the $(\vec{k}+ \vec{s})\uparrow$ and $(-\vec{k}+\vec{s})\downarrow$ states forming a pair with drift momentum
$2\hbar\vec{s}$ exceeds the pair binding energy $2|\Delta_{\vec{k}}|$. An empirical model along these lines  gives a
reasonably good account of the available experimental data, although further experiments on the same crystal at
closely spaced doping levels may be worthwhile. The model gives effects of the same magnitude as  Yip and Sauls
predicted at low T, but they persist to higher temperatures. It could also be checked experimentally by measuring zinc
doped crystals where the NLME is predicted to be shifted to higher fields because of increased pair-breaking in zero
field\cite{loramzn,bonn}. Extensions of the present approach to higher Doppler shifts ($d$ greater than $k_{B}T$)
could also be relevant for understanding some properties of cuprate superconductors at higher magnetic fields, for
example the penetration depth\cite{sonier} and the thermal conductivity in the vortex state\cite{behnia}. However its
relation to the usual theory involving the Doppler shift\cite{volovik,kubert} needs a more careful study because both
pictures predict similar numbers and types of quasi-particles at high fields and the Doppler shift approach does
appear to agree with recent thermal conductivity experiments\cite{chiao}.

\section{THE EMPIRICAL MODEL}
The total supercurrent density $\vec{j}$ in a superconductor is usually expressed as the sum of the contribution
$\vec{j}_{s}$ from Cooper pairs with a drift velocity $\vec{v}_{s}$ and the contribution from the backflow of
quasiparticles $\vec{j}_{qp}$ \cite{scalapino}. For a cylindrical Fermi surface(FS), with quadratic in-plane energy
dispersion, and $\vec{v}_{s}$ in the conducting planes,
 $\vec{j}_{s} =
ne\vec{v}_{s}$ where $n$ is the total number of carriers per unit volume.
 As far as I know,
 published theoretical treatments of the NLME only consider
non-linearity in the quasiparticle term, and implicitly assume that the gap parameter $\Delta_{\vec{k}}$ retains its
usual $d$-wave form. For example \cite{scalapino}, for a cylindrical FS, Equations \ref{eqn1} and \ref{eqn2} together
with the relation $\vec{j}$ = $\vec{j}_{s}$ + $\vec{j}_{qp}$, lead to all the non-linear effects proposed until now.
\begin{eqnarray}
 \label{eqn1}\nonumber
j_{qp}&=&\frac{4en}{mv_{F}^{2}}\int_{0}^{\infty}\frac{d\epsilon}{2\pi}\times\\
&&\int_{0}^{2\pi}d\theta
 v_{F}\cos\theta \, f(\sqrt{\epsilon^{2}+ \Delta^{2}(\theta)} +
 mv_{F}v_{s}\cos\theta)
\end{eqnarray}

In this Equation, $f$ is the Fermi function, $e$ the electronic charge, $v_{F}$ the Fermi velocity, $\epsilon$ and $m$
the energy and effective mass of the carriers respectively.  The superfluid velocity $\vec{v}_{s}$ is directed along
the crystallographic $\vec{x}$ axis and the angular dependent superconducting energy gap is given by:

\begin{equation}
\label{eqn2} |\Delta_{\vec{k}}(T)|=\Delta(T,\theta) = \Delta(T)|\cos2\theta|
\end{equation}

\begin{figure}
\centerline{\psfig{figure=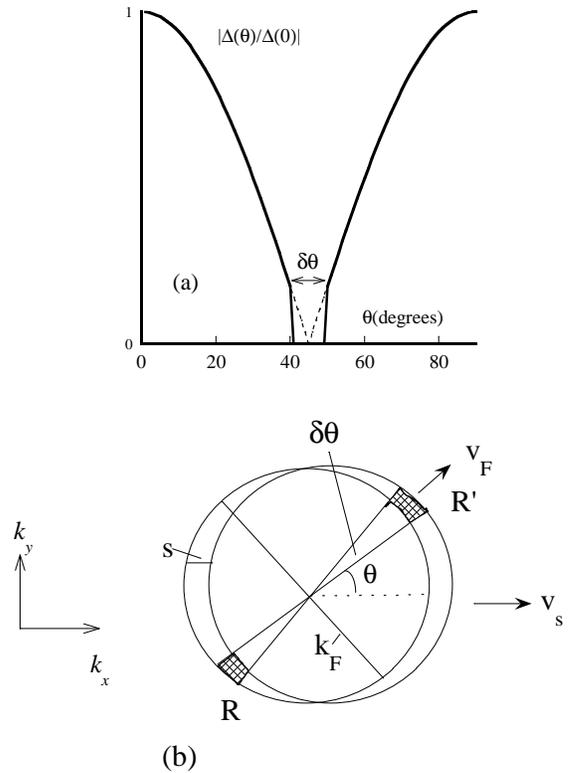,width=8cm}} \vskip 0cm \caption{(a) Suggested angular variation of
$|\Delta(\theta)|$ in the presence of a finite supercurrent (solid line). (b) Section of a cylindrical Fermi surface
showing the angular regions of width $\delta\theta$ where pairing is suppressed. The loss in current density from
Cooper pairs in the angular range $\delta\theta$  is equal to 2e$v_{F}\cos\theta$ (the current change on moving one
carrier from R' to R) multiplied by the number of carriers per unit volume in the shaded areas R or R'. See text.}
\label{fig1}
\end{figure}

At low T when the Doppler energy shift of one electron in a Cooper pair, $d =
\frac{1}{2}m[(\vec{v}_{F}+\vec{v}_{s})^{2}- (\vec{v}_{F})^{2}] = m\vec{v}_{F}.\vec{v}_{s}\gg k_{B}T$, Equations
\ref{eqn1} and \ref{eqn2} give the non-analytic variation $\lambda  \propto |v_{s}|$, i.e. $\lambda \propto |H|$ that
was predicted by Yip and Sauls\cite{Yip}.  In the opposite limit, $d\ll k_{B}T$, $\delta\lambda/\lambda$ is
expected\cite{scalapino} to vary as $H^{2}/T$. Here I suggest that Equation \ref{eqn2} does not hold exactly except
when $v_{s}\rightarrow 0$.  Namely, as sketched in Figure 1(a), in the parts of the FS where $|d| \geq
|\Delta_{\vec{k}}(T)|$, $\Delta (T) = 0$, the Cooper pairs are actually broken and do not contribute to the
supercurrent\cite{helpful}. This differs from the usual picture,  shown in Figure 2, where the FS is still displaced
near the nodes and the pair current there is completely compensated by the back-flow when (a)  $d \gg k_{B}T$ and
$d\geq\Delta(\theta)$, giving strong non-linearity\cite{Yip} in $\lambda$, or (b)  $d\ll k_{B}T$ and $k_{B}T \gtrsim
3\Delta(\theta)$ and the non-linearity in $\lambda$ is weak\cite{Yip}. Pair-breaking is plausible physically because
when  $|d| \geq |\Delta_{\vec{k}}(T)|$ the energy difference between the two electron states which form a Cooper pair
is larger than the pair binding energy, $2|\Delta_{\vec{k}}(T)|$, and in an s-wave superconductor, where
$\Delta_{\vec{k}}$ is uniform, this condition corresponds to the critical (depairing) current.

 To some extent it can be justified within weak-coupling BCS\cite{BCS} theory as follows. Using the standard d-wave
 pairing interaction \cite{maki}, $V_{kk'}= -Vcos2\theta_{k} cos2\theta_{k'}$, the BCS Equations\cite{BCS} 3.15 and
 3.16  can be used to compute the increase in free energy F when $\Delta(\theta,T)$ is set equal to zero in a small
 angular region $\delta\theta$ near the four nodes.
For the stationary superfluid this costs an energy:

\begin{equation}
\label{eqn3}
  \delta F=\beta \Delta(T,\pi /2)^{2}\delta \theta^{3}N(0)
\end{equation}
where N(0) is the density of states at the FS for one spin direction, and $\beta$ is a   weakly T dependent constant,
which ranges from 1.49 at $T/T_{c}$ = 0.16, to 1.19 at $T/T_{c}$ = 0.67. If only two nodes are affected (e.g. for
current flow in the nodal directions) then $\beta$ is reduced by a factor of two.  For the moving superfluid, there
are no Cooper pairs in this small angular region, so the kinetic energy $K$ is reduced, and in fact $\delta K$ varies
linearly with $\delta\theta$. However there is also a magnetic self energy term (M) arising from the magnetic field
produced by the supercurrents which has the form \cite{waldram} $\vec{j}.\vec{A}$, where $\vec{A}$ is the magnetic
vector potential.  In the London gauge \cite{waldram} $\vec{A} = m\vec{v}_{s}/e$ and I argue below that $\vec{j}_{qp}$
is not altered by pair-breaking at the nodes.  Hence there is a gain in magnetic energy of the form $m \delta
\vec{j}_{s}.\vec{v}_{s}/e$ arising from the absence of Cooper pairs near the nodes. Referring to Figure 1(b), $\delta
\vec{j}_{s}.\vec{A}$ is given by $2e \vec{v}_{F}.\vec{A}$, multiplied by the total number of carriers in the region R,
namely $s\cos\theta k_{F}\delta\theta n/\pi k_{F}^{2}$ = $nv_{s}\delta\theta \cos\theta/\pi v_{F}$. Hence for one pair
of opposite nodes:

\begin{figure}
\centerline{\psfig{figure=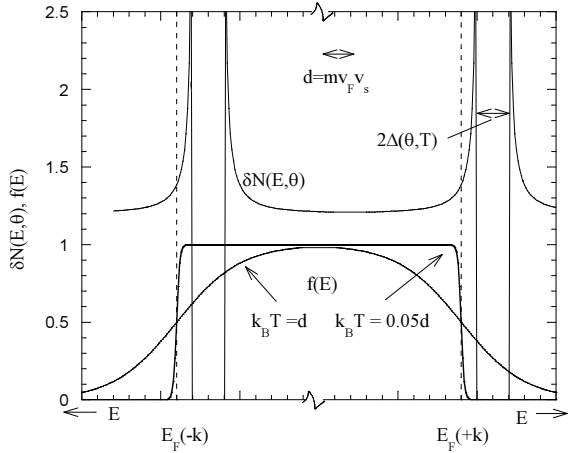,width=8cm}} \vskip 0cm
 \caption{Partial density of states(DOS) for Doppler
shifted Bogoliubov quasi-particles at an angle $\theta$. At low T, e.g. $k_{B}T = 0.05d$ where $d$ is the Doppler
shift, the backflow of quasi-particles in angular regions where the superconducting gap $\Delta(\theta,T)< d$ exactly
compensates the flow of Cooper pairs arising from the shifted DOS in those regions \protect\cite{Yip}. At high T, e.g.
$k_{B}T = d$, this does not happen.} \label{fig2}
\end{figure}

\begin{equation}
\label{eqn4}
 \delta M = \frac{2}{\pi}n m v_{s}^{2}\delta\theta \cos^{2}\theta
\end{equation}
By integrating over the whole of the FS, from $\theta$ = 0 to $\pi$, it can be seen that $M = n m v_{s}^{2}$ = 2$K$.
Minimising the sum of $\delta F$ and $\delta (M+K) $  with respect to $\delta \theta$ gives:

\begin{equation}
\label{eqn5}
    \delta\theta = 0.64\frac{mv_{F}v_{s}}{\Delta (T)}
\end{equation}
when $\vec{v}_{s}$ is along a nodal direction.  $\delta \theta$ is a factor of $\sqrt{2}$ smaller when
$\vec{v}_{s}\parallel \vec{x}$. This expression has  the same form and the same anisotropy as the qualitative argument
given above, the presence of the extra factor 0.64 worsens the agreement with experiment slightly and has been omitted
in subsequent discussion.  This may be an indication that there is another pair-breaking mechanism or another term in
the free energy varying linearly with $\delta\theta$, not covered by the above treatment, or that the numbers are
changed slightly by strong coupling effects since it is known that for many cuprates  $\Delta(0)\approx 3k_{B}T_{c}$
rather than the weak coupling d-wave value, $\Delta(0) = 2.14k_{B}T_{c}$.

The effect of the broken pairs on the penetration depth can be estimated in a similar way. A crucial question is
whether the carriers produced by the broken pairs will have zero drift velocity in the laboratory frame or whether
they will be in equilibrium with the pair current and the Bogoliubov quasiparticles on the remainder of the FS. It
seems that only in the latter case will the present model account for the behaviour of $\lambda$(H). From Figure 2 it
can be seen that in both cases (a) when $k_{B}T \ll d$ and (b) when $k_{B}T \gg d$ and $k_{B}T \gg\Delta(T,\theta)$,
the pair supercurrent in that angular region will be balanced by the backflow, to give almost zero net supercurrent,
therefore producing stationary carriers by pair-breaking will have almost no effect on $\lambda$. This is confirmed by
numerical calculations based on Equation \ref{eqn1}, namely, if the contributions to $j_{qp}$ and $j_{s}$ from the
angular region $\delta\theta$ are simply omitted there is almost no change in $\lambda$ for $k_{B}T \gg d$. On the
other hand if the carriers from broken pairs have the same drift velocity as the Bogoliubov quasiparticles then
$j_{qp}$ can be calculated from Equation \ref{eqn1}, simply by setting $\Delta(\theta) = 0$ near the nodes. For
$k_{B}T \gg d$, $\delta j_{qp}$ only varies weakly with $d$, as $d^{2}$, while $\delta j_{s}$ varies as $d^{1}$.
Assuming then that the carriers do have the same drift velocity as the Bogoliubov quasiparticles, the loss in
supercurrent associated with the broken pairs is given by $\delta j_{s} = 2nev_{s}\delta\theta \cos^{2}\theta/\pi$ for
a pair of nodes. This is the same formula used above when calculating the magnetic self energy of the supercurrents.
Hence from this formula, $\delta n/n = \delta j_{s}/j_{s} = -2\delta\theta/\pi$ for both $\vec{v}_{s}$ parallel to the
nodal directions (2 nodes with $\cos\theta = 1$) and to the anti-nodal directions (4 nodes with $\cos\theta =
1/\sqrt{2}$).

The penetration depth $\lambda(T) = \sqrt{mc^{2}/4\pi e^{2}(n-n_{qp})}$, and neglecting non-linear effects on $n_{qp}
= |\vec{j_{qp}}/e\vec{v}_{s}|$ for the reasons given above, this  changes by:

\begin{equation}
\label{eqn6}
    \frac{\delta\lambda}{\lambda} = -\frac{1}{2}\frac{\delta n}{(n-n_{qp})}
\end{equation}

Using the relation $ H = 4\pi I/c$ between the surface field ($H$) and the surface current $I$ per unit length yields
$H = 4\pi j\lambda/c$, which together with the above expressions for $\lambda$ and $\delta\theta$ (but omitting the
factor of 0.64 in Equation \ref{eqn5}) gives:

\begin{eqnarray}\nonumber
\label{eqn7} \frac{\delta \lambda (H)}{\lambda(T)}&=& \frac{\gamma (T) d }{\sqrt{2}\pi\Delta(T)} = \frac{ev_{F}\mid
H\mid \lambda(T)\gamma (T)}{c\sqrt{2} \pi \Delta(T)}\\&=& \frac{\pi\mid H\mid \xi(T)\lambda(T)\gamma
(T)}{\sqrt{2}\Phi_{0}}
 = \alpha\gamma(T)\frac{\mid H \mid}{H_{c}(T)}
\end{eqnarray}

This expression refers to the case where the current flow is at 45$^{o}$ to the nodes. According to the present
treatment the effect will be a factor of $\sqrt{2}$ larger when the current flows in the nodal directions, as in the
work of Yip and Sauls\cite{Yip}. $\Phi_{0}$ is the flux quantum for pairs, $\xi(T) = \hbar v_{F}/\pi\Delta(T)$ is a
measure of the in-plane coherence length for a $d$-wave superconductor and $\gamma(T)
\equiv\lambda(T)^{2}/\lambda(0)^{2}$. $H_{c}(T)$ is the thermodynamic critical field, if the standard Ginzburg-Landau
expression for an s-wave superconductor, $H_{c}(T)= \Phi_{0}/2\sqrt{2}\pi\lambda(T)\xi(T)$ is used, the constant
$\alpha$ = 0.25. Estimates using parameters for a cylindrical FS give $\alpha$ = 0.16 for the weak coupling $d$-wave
case, (for which $\Delta(0) = 2.14 k_{B}T_{c}$ and $H_{c}(0)^{2}/8\pi = 0.125N(0)\Delta(0)^{2}$) and $\alpha$= 0.129
for the more realistic case $\Delta(0) = 3k_{B}T_{c}$ and \cite{lorampriv} $H_{c}(0)^{2}/8\pi =
0.089N(0)\Delta(0)^{2}$. Equation 7 has  similar form and similar magnitude to the Yip-Sauls result at low T, but it
has a different origin, in the present work it arises because the superconducting gap parameter has been suppressed
near the nodes, where it no longer obeys Equation \ref{eqn2}. The extra factor $\lambda(T)^{2}/\lambda(0)^{2}$ results
from the assumption that $j_{qp}$, i.e. $n_{qp}$, is not changed at all by the presence of pair-breaking. If it were
changed in the same ratio as $n$, i.e. if $\delta n_{qp}/n_{qp}$ =   $\delta n/n$, then the
$\lambda(T)^{2}/\lambda(0)^{2}$ factor would be absent and the temperature dependencies would be in much better
agreement with experiment. But there does not seem to be any theoretical justification for such an "ad hoc"
hypothesis. The $|H|$ behaviour in Equation 7 will be cut off at low fields by the presence of pair-breaking from
other sources, for example at low T from impurities in the CuO$_{2}$ planes and at higher T by inelastic scattering
processes.  This effect can be taken into account by replacing $|H|$ by the empirical expression $\sqrt{H^{2}+
H_{0}^{2}}-H_{0}$, where $H_{0}$ is a measure of the pair-breaking effects in the absence of an applied field.

\begin{figure}
\centerline{\psfig{figure=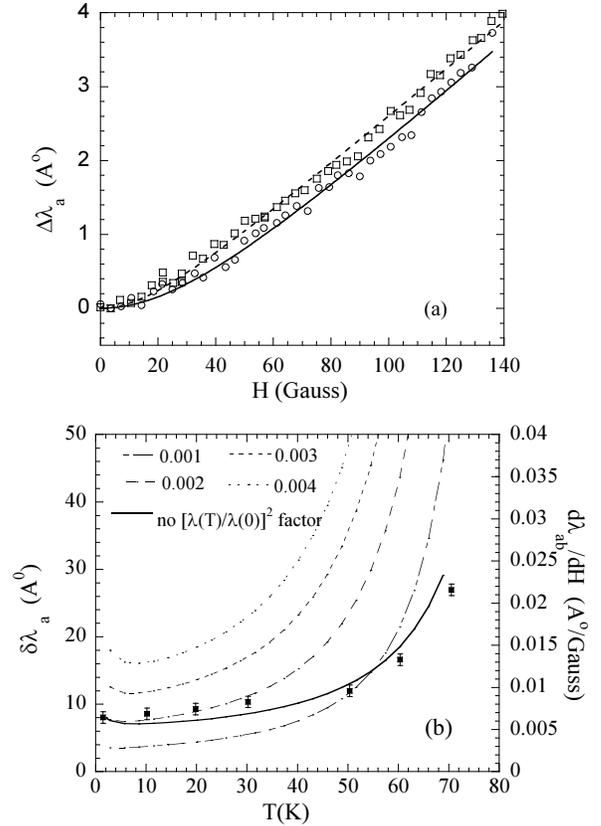,width=8cm}}
\vskip 0cm
\caption{(a) Field dependence of the $a$-axis penetration
depth of YBa$_{2}$Cu$_{3}$O$_{6.95}$ measured\protect\cite{bidinosti} for a detwinned crystal at 4.2 K (circles) and 7
K (squares). The  dashed and solid lines are fits to the data using Equation 7 together with cut-off fields $H_{0}$ of
23 and 42 Gauss respectively. (b) Measured \protect\cite{carrington} temperature dependence of d$\lambda_{ab}$/d$H$
(squares) for a YBa$_{2}$Cu$_{3}$O$_{6.95}$ crystal with $T_{c}$ = 91.4 K. The dashed lines show the behaviour
calculated according to Equation 7 for various magnetic fields corresponding to $d/\sqrt{2}\Delta(\frac{\pi}{2},T=0)$
= 0.001 to 0.004.  The solid line shows the effect of omitting the $\lambda(T)^{2}/\lambda(0)^{2}$ factor, see text.}
\label{fig3}
\end{figure}

\section{COMPARISON WITH EXPERIMENT}
Experimental data\cite{bidinosti} for the field dependence of $\lambda_{a}$ for a detwinned
YBa$_{2}$Cu$_{3}$O$_{6.95}$ crystal are shown in Figure 3a, together with fits to Equation 7, including the cut-off
parameter $H_{0}$. The linear terms in these fits correspond to d$\lambda_{a}$/d$H$ values of 0.033 $\pm$ 0.001
A$^{o}$/Gauss  at both 4.2 and 7 K.  In contrast to the Yip-Sauls result, the slope calculated from Equation 6 is
T-independent at these temperatures. Using the experimental value\cite{bidinosti} $\lambda_{a}(T=0)$ = 1600 \AA,
leaves one free parameter $\xi_{a}(0)$ which has to be equal to 16.4 \AA to account for the data in Figure 2(a).
This value is quite compatible with $\xi_{a-b}(0)$ = 11 \AA obtained previously by analysing\cite{philmag} the
specific heat above $T_{c}$ in terms of Gaussian fluctuations bearing in mind that $\xi_{a-b}(0)$ contains a
contribution from the Cu-O chains, depends on the FS average of $v_{F}$ and is related to the RMS average of
$\Delta(0)$. Taking $2\Delta(0) = 6k_{B}T_{c}$, the value of $\xi_{a}(0)$ obtained in the present work corresponds
to an acceptable value for $v_{F} = 1.9 \times 10^{7}$ cm/s in the nodal directions. Using Equation 6 and the same values
of $\Delta(0)$ and $\lambda_{a}(0)$, the value of $H_{0}$ = 30 $\pm$ 10 Gauss,  obtained from the fits, gives a very
small pair-breaking energy of 0.5 $\pm$ 0.16 K. This compares well with typical values of the pair-breaking
temperature $T^*$ $\leq$ 1 K obtained from the $T$ dependence of $\lambda$ in zero field\cite{carrington,bonn}.

The measured $T$ dependence\cite{carrington} of d$\lambda_{ab}$/d$H$ is shown in Figure 3b. The fit to Equation 7 using
the $d$-wave behaviour of $\Delta(T)$ and $\lambda(T)$, calculated in the weak-coupling limit for a cylindrical FS,
gives the correct general behaviour with some deviations at high $T$. Exact agreement is not to be expected in view of
the fact that the weak-coupling ratio  $2\Delta(0)$ = $4.28k_{B}T_{c}$ is smaller than typical experimental values,
$2\Delta(0)$ = $6k_{B}T_{c}$. However as also shown in Figure 3b, omission of the $\lambda(T)^{2}/\lambda(0)^{2}$
factor gives much better agreement with experiment.

\begin{figure}
\centerline{\psfig{figure=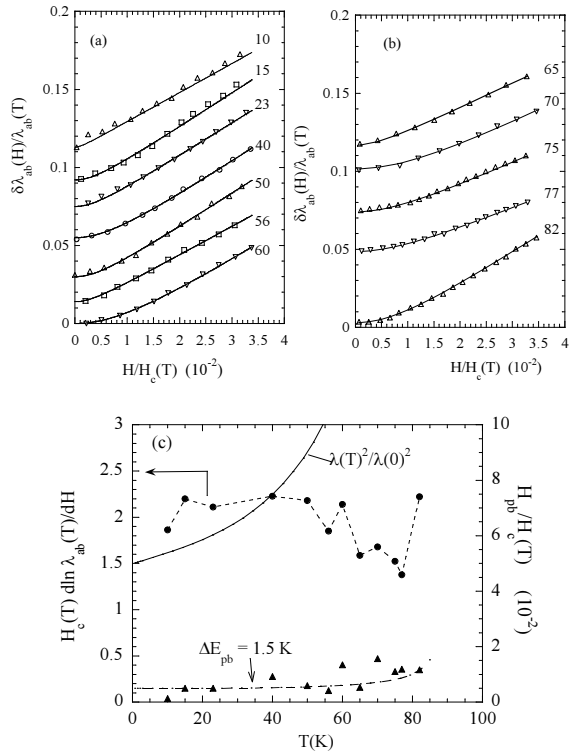,width=8cm}} \vskip 0cm
\caption{(a)
and (b) Experimental data for the in-plane penetration depth of a Bi:2212
crystal\protect\cite{Maedarev} as a function of the applied field,  $H
\parallel c$, normalised to the thermodynamic critical field $H_{c}(T)$
at various  temperatures. The solid lines show fits to Equation 7. (c)
Values of the parameters $\alpha$(left-hand scale) and $H_{0}$/$H_{c}(T)$
(right- hand scale) determined from these fits, omitting the
$\lambda(T)^{2}/\lambda(0)^{2}$ factor, which as shown by the solid line
gives the wrong $T$ dependence. The dashed line shows the $T$ dependence
expected if the pair-breaking parameter $H_{0}$ is constant.}
\label{fig4}
\end{figure}

Experimental data for $\delta\lambda_{ab}(H,T)/\lambda_{ab}(0,T)$ versus
$H/H_{c}(T)$ from the work of Maeda and colleagues\cite{Maeda,Maedarev}
for Bi:2212 crystals are shown in Figure 4(a).  $H_{c}(T)$ was derived
experimentally\cite{Maeda,Maedarev} for the same crystals by assuming
that the lowest field $H^{*}$, at which some hysteresis could be detected
in field sweeps, was a measure of the lower critical field $H_{c1}(T)$.
Then $H_{c}(T)$ was obtained by taking a plausible value $\kappa = 100$
for the Ginzburg-Landau parameter.  It can be seen that the data can be
fitted to Equation 7 at all reduced temperatures ($t = T/T_{c}$) studied.
However as shown in Figure 4(b) the values of the slopes  are essentially
constant rather than increasing as $\lambda(T)^{2}/\lambda(0)^{2}$ as
predicted by Equation 7. Again this extra factor of
$\lambda(T)^{2}/\lambda(0)^{2}$ is a weak point of the present model.
Even without the latter factor the magnitude of the slopes, i.e., $\alpha
$, are  about a factor of 10 larger than that calculated above. It was
suggested\cite{Maedarev} that this was mainly because the effect of the
geometrical barrier\cite{zeldov} leads to $H_{c1}(T)$ and hence
$H_{c}(T)$ being overestimated.  The crystals used were slightly
overdoped with $T_{c}$ values of  90 K.  Although it is known that the
condensation energy of Bi:2212 can be reduced by a factor of 2-5 by the
presence of the normal state pseudogap\cite{jwlhouston}, even for such
high values of $T_{c}$, this effect is probably taken into account by
using experimentally determined values of $H_{c}(T)$.   The pair-breaking
parameter $H_{0}(T)/H_{c}(T)$ obtained from these data is also plotted in
Figure 4 (b). $H_{0}$ is essentially constant in the temperature range
studied, showing that inelastic scattering processes are not strong
enough to cause pair-breaking.

\section{SUMMARY AND CONCLUSIONS}
I have argued that the non-linear Meissner effect in the cuprates can be
understood in terms of an empirical model in which the flow of a finite
supercurrent destroys superconducting order in a small angular range near
the $d$-wave nodes.  This model accounts for some of the experimental
data and would mask the unusual temperature dependence predicted by Yip
and Sauls. The general viewpoint outlined here is reminiscent of
published theoretical work suggesting that application of high fields
could actually alter the symmetry of the order parameter, for example by
introducing a $d + id$ \cite{balatsky} or a $d + id_{xy}$
component\cite{laughlin}.  However for experiments at low fields it seems
that the empirical pair-breaking picture described here could be
adequate. The present approach, which suggests that the pair-breaking is
an intrinsic property, still predicts anisotropic effects for which there
is presently no experimental support. Although in this regard one should
note that there are potential complications in separating Cu-O$_{2}$
plane  and Cu-O chain contributions to the penetration depth in YBCO and
that in the present model  the details of any non-linear effects in the
magnetisation will be altered by the different behaviour of the free
energy. The present work supports the empirical
conclusion\cite{Maeda,Maedarev,Maedael} that even though the high
temperature NLME is not predicted by the original Yip-Sauls theory, it is
nevertheless a characteristic feature of {\lq}non s-wave{\rq} superconductivity.

There are indications that measurements of the non-linear Meissner effect as a function of hole concentration could be
used as a means of investigating the effect of the normal state gap on the superconducting properties and for
investigating pair-breaking mechanisms in zero applied field.

\section{ACKNOWLEDGEMENTS}
I am grateful to A. Carrington and J.W. Loram for several helpful suggestions and for critically reading the
manuscript. Helpful comments from R.W. Giannetta, W.N. Hardy, A. Maeda, K. Maki, D. Scalapino, Tao Xiang and J.R.
Waldram are also gratefully acknowledged.


\begin{references}
\bibitem{helpful} It may be helpful to remind the reader that in BCS theory, pairs with arbitrary small values of
$\Delta_{\vec{k}}$ still contribute to the pair supercurrent  because the magnitudes of the coefficients $u_{\vec{k}}$
and $v_{\vec{k}}$ in the BCS wave-function are always $1/\sqrt{2}$ for $|\vec{k}|= k_{F}$, but if
$|\Delta_{\vec{k}}|\ll k_{B}T$ this will be completely compensated by the back-flow of Bogoliubov quasi-particles. In
this paper the term pair-breaking or broken pairs means that both $u_{\vec{k}}$ and
$v_{\vec{k}}$ are zero in a finite angular region near the nodes.
\end{references}
\end{document}